# CONGESTION AWARE SPRAY AND WAIT PROTOCOL: A CONGESTION CONTROL MECHANISM FOR THE VEHICULAR DELAY TOLERANT NETWORK

Chuka Oham and Milena Radenkovic

School of computer science, University of Nottingham, United Kingdom

## *ABSTRACT*

*In the last few years, the Vehicular Ad-hoc Network (VANET) has come to be an important area of research. Significant research has been conducted to improve the performance of VANETS. One output of further research conducted on VANET is the Vehicular Delay Tolerant Network (VDTN). It is an application of the mobile DTN where nodes relay messages in the network using a store-carry-forward approach. Due to its high mobility, it suffers frequent disconnections and also congestions at nodes which leads to message drops. To minimize the rate of message drops and so optimize the probability of message delivery so that drivers are increasingly aware of the situation of the road, we propose a congestion control mechanism: Congestion Aware Spray and Wait (CASaW) protocol in this work so as to optimize the rate of message delivery to its destination and so increase the awareness of drivers in the vehicular environment thereby improve road safety. The results have shown that our proposition performed better than other classical VDTN protocols in terms of message delivery probability and rate of packet drops performance measures. We used the Opportunistic Networking Environment (ONE) simulator to implement the classical VDTN protocols: the PROPHET protocol, the Epidemic protocol, the MaxProp protocol and the Spray and Wait Protocol. The simulation scenarios shows a better performance for the congestion control mechanism we propose as it maintains a good message delivery rate as well as minimize the rate of packet losses thereby optimizing the chances of messages getting to their destinations and so improve road safety.*

## 1. INTRODUCTION

Vehicular Adhoc Networks (VANETs) are a special class of the Mobile Adhoc Networks with some distinguishing characteristics [1]. Unlike the Mobile Adhoc Networks, it is known for its predictive mobility and high mobility which leads to frequent disconnections in areas of low traffic and nodal congestions in regions of high traffic. These factors impact on the network performance and so researchers proposed the extension of the Delay Tolerant Networking (DTN) paradigm to the Vehicular Adhoc Networks to overcome the limitations [2].

Delay Tolerant Networks (DTN) are networks that enable communication where connectivity issues such as delays, intermittent connectivity, high error rates etc. exists and also in regions where end-to-end connectivity may not exist. Communications exists where network nodes are custodians of the message been transmitted and so forwards messages when opportunity arise using a store-carry-forward approach. The extension of the DTN to the VANET gave rise to the Vehicular Delay Tolerant Networks (VDTN). With this approach, a vehicle stores a message in its buffer and carries it along with it and when an opportunity to forward the message to another node arises, it forwards it.

Vehicular Delay Tolerant Network (VDTN) is an application of mobile DTN. It is a family of opportunistic, autonomous and self-organized networking area that arose from the implementation of wireless communication where network interferences and disruptions are a mainstay [3]. In this paper, we consider the use of the VDTN protocol to provide a congestion control solution to make network nodes congestion aware, optimize the probability of message delivery and minimize the rate of message loss in the VDTN and possibly increase the chances of improving the awareness of drivers on the road.

In order to cope with frequent disconnections, network nodes store data in their buffers for a period of time awaiting an opportunity to forward the data to intermediate nodes in the network or to the final destination. This means that the buffer capacity of nodes in the network may likely impact on the performance of the in terms of message delivery probability and the rate of packet loss. Using the Helsinki map based model of the ONE simulator, we run simulations to analyze the efficiency of our congestion aware protocol and the four generic VDTN protocols when the buffer capacities of nodes are varied and then comparatively evaluate their performances in terms of probability of message delivery and the rate of packet loss.

The rest of this paper is organized thus: section 2 describes related works on congestion in the VDTN and highlights our contributions, section 3 describes the generic VDTN routing protocols, section 4 describes our proposed congestion aware protocol, section 5 presents simulation scenarios and an analysis of results obtained while section 6 is the conclusion of the paper and recommendations for future work.

## 2. RELATED WORKS

Congestion occurs at any given time when the sum of demand exceeds available capacity. In the DTN, earlier works in congestion control focused primarily on the challenge of reducing the delivery latency and the associated cost with the underlying assumption of an unlimited storage capacity [4]. Congestion control is an important feature in the DTN that directly impacts on the network's performance. Network congestion may either cause significant network delay or loss of message which in the VDTN may lead to loss of life. The control of congestion in the VDTN is therefore necessary when the buffer of nodes are over-utilized.

Y. An et al [5] divided the congestion control mechanism into the proactive and reactive policies. These are buffer management policies used to control congestion in the network.

- ***Proactive policy***: With this policy, an admission control is applied at the beginning to avoid congestion network. This policy mechanism adopts an economics model. It compares the reception and dissemination of messages to the risk investment. When a message arrives at a node, it decides either to accept or reject the message in accordance to the risk value of receiving and storing the message. The value of risk is dependent on the buffer space of the node, the data input rate, the residual TTL of the message etc.

- ***Reactive policy***: This policy mechanism is used to perform some response action when congestion has already occurred in the network. L. Amornsin et al [6] described some response actions when congestion occurs.

**\*** Last in First Drop: A simple dropping scheme where nodes reject message once buffer capacity is fully utilized.
**\*** Oldest Drop: This is based on the time-to-live (TTL) of a message in a node. The message with the least TTL is dropped first.
**\*** Youngest drop: Proposed by Lindgren et al. In this case, the message with the largest TTL is dropped.

Several congestion control mechanisms have been proposed by researchers to minimize congestion in the VDTN. Zhang et al [7] proposed a proactive cross-layer congestion control through dynamic transmit power control. The transmit power control is used to optimize energy consumption as well as the point-to-point connectivity. Kirfa et al [8] discovered that traditional buffer management policies implemented to minimize congestion in the VDTN are suboptimal. They therefore used the theory of encounter based message dissemination to propose an optimal buffer management policy based on the knowledge of the network. They derived the global knowledge of the network using a distributed algorithm that uses statistical learning. Radenkovic et al [4] Proposed an for congestion control in social opportunistic networks. They suggested a combination of routing and congestion avoidance that implements heuristics to deductively derive shorter paths to destinations from social information.

As opposed to existing approaches, we will in this work, design a congestion control mechanism that implements a buffer-check-status feature that will prior to forwarding message to another node in the network, check that the receiving node has sufficient buffer space to accommodate the new message. If space is sufficient, the message is forwarded to the node else the message will not be forwarded to it except the node is the destination node. If this is the case, a buffer policy will be applied where the oldest message in the buffer of the node is dropped to make room for the new message.

## 3. VDTN ROUTING PROTOCOLS

In the VDTN, the mobility of nodes enables opportunistic communications. Data is disseminated using the tore-carry-forward DTN approach. Using this approach, data is replicated by multiple nodes until it gets to the destination. Vasco et al [2] identified four generic DTN protocols that have been implemented in the VDTN. These protocols are the epidemic protocol, the MaxProp protocol, the PROPHET protocol and the Spray and wait protocol.

### 1.1. Epidemic Protocol

Proposed by Vahdat et al [9]. It is a flooding based routing technique where a message is replicated across all nodes in the network to increase the probability of the message getting to the destination. This flooding based technique wastes network resources because messages are flooded throughout the entire network to get to one destination. This creates contention for buffer space and so, leads to congestion in the network and contention for transmission time.

### 1.2. MaxProp Protocol

Proposed by Burges et al to address scenarios where there is limited transfer time or storage in the network. It uses hop count in packets as a measure of network fairness to mitigate bias towards short distances. It also uses acknowledgements (ACKs) which are propagated throughout the network to remove stale data from the buffer [10].

### 1.3. PROPHET Protocol

Proposed by Lindgren et al [11]. This approach uses an estimation of delivery to determine performance measures such as the delivery probability or delivery delay relative to the successful delivery of a message. It works on the premise that the mobility of nodes are not truly random. It uses a metric called delivery predictability to estimate the probability of a node delivering a message to another node.

### 1.4. Spray and Wait Protocol

Proposed as an overlay of the flooding based strategy by Spyropolous et al [12]. It is an efficient routing protocol that performs fewer transmissions than all flooding based schemes. it generates low contention under high traffic situations. Y. Shao et al [13] said that it was introduced to reduce the wasteful flooding of redundant messages in the network. Compared to the epidemic protocol, it limits the number of disseminated copies of same message to a constant L. Spray and wait occurs in two phases [14]. The spray and the wait phase. In the spray phase, for every message that originates from a source L, copies of that message is forwarded by the source and the other nodes receiving the message to a total of L distinct relays. In the wait phase, all L nodes with a copy of the message performs direct transmission to the destination.

Several studies and evaluations have been conducted to ascertain the most effective VDTN protocols. Spyropolous et al [12] showed from simulations that the spray and wait protocol outperforms other VDTN protocols with regards to average message delay and the number of transmissions per message delivered. To support the claim of Spyropolous et al, we conducted multiple simulation runs to comparatively evaluate the performance of the VDTN protocols and observed from the result that the spray and wait protocol performed significantly better than other protocols. In line with this observation, we will design and implement a congestion aware spray and wait protocol for the VDTN to try to minimize message loss and optimize message delivery.

## 4. CONGESTION AWARE SPRAY AND WAIT PROTOCOL

The routing protocols in the opportunistic networking environment (ONE) simulator lacks the implementation of a congestion control mechanism to help minimize the rate of message loss in the VDTN. Therefore we will implement a new congestion aware algorithm to control congestion in the vehicular network so as to try to increase the likelihood of messages getting to their destination and also reduce the risk on the road by minimizing the rate of packet losses. To achieve this, we implemented three states of congestion. *The buffer is filled up state, the buffer is occupied state and the buffer is empty state.* Nodes are said to be in the *buffer is filled up state* when they can no longer receive messages because of the state of the buffer. In this state, the free buffer space is zero *(fbs=0)*. To achieve fewer drops when this occurs, we implemented a *buffer-check-status* feature for nodes in the network so that they are apprised of the buffer states of other nodes in the network prior to making forwarding decisions. In *the buffer is occupied state*, nodes in the network are said to be occupied but not totally congested. To prevent congestion and achieve fewer drops in this state, we allowed nodes in the network to keep receiving messages until a point where the size of the incoming message becomes bigger than the size of the free buffer space (IncomingMsgSize > fBS) or a point where the buffer becomes totally filled up. If the incoming message size is bigger than the free buffer size and the node is not the destination node, the node is ignored and the message is forwarded to another node but if the node is the destination node, we applied our buffer drop policy to accommodate the incoming message. If however, the node receives message to a point where the buffer becomes totally filled up, the next incoming message will be received only if the node is the destination node and after the buffer policy has being applied. *In the buffer is empty state*, the buffers of nodes are empty so we allowed the reception of messages until either the incoming message size becomes bigger than the free buffer space of the node or the buffer of a node becomes totally filled up. When either of this occurs, the reception of messages and the application of buffer policies follow same process as when the nodes are in *the buffer is occupied state*.

International Journal of Computer Science & Information Technology (IJCSIT) Vol 7, No 6, December 2015

## 1.1.Key Design Features of the Congestion Aware Spray and Wait (CASaW)

The figure (1.0) below shows the interaction of nodes in the VDTN implementing our proposed Congestion Aware Spray and Wait (CASAW) algorithm. Our CASaW introduces a new approach to the existing spray and wait methodology. Our approach compared to the existing 2-phased spray and wait occurs in 3-phases. The *Check phase*, the *Spray phase* and the *Wait phase*. In the *Check phase*, the buffer space of nodes are checked to see if they can contain the message. In the *Spray phase*, messages are sent to nodes that can only receive the message except the node is the destination and is congested. In this case, we applied a buffer policy of deleting the oldest message in the buffer to make room for the new message. In the *Wait phase*, the nodes perform direct delivery to the destination if the destination is not identified in the *Spray phase*.

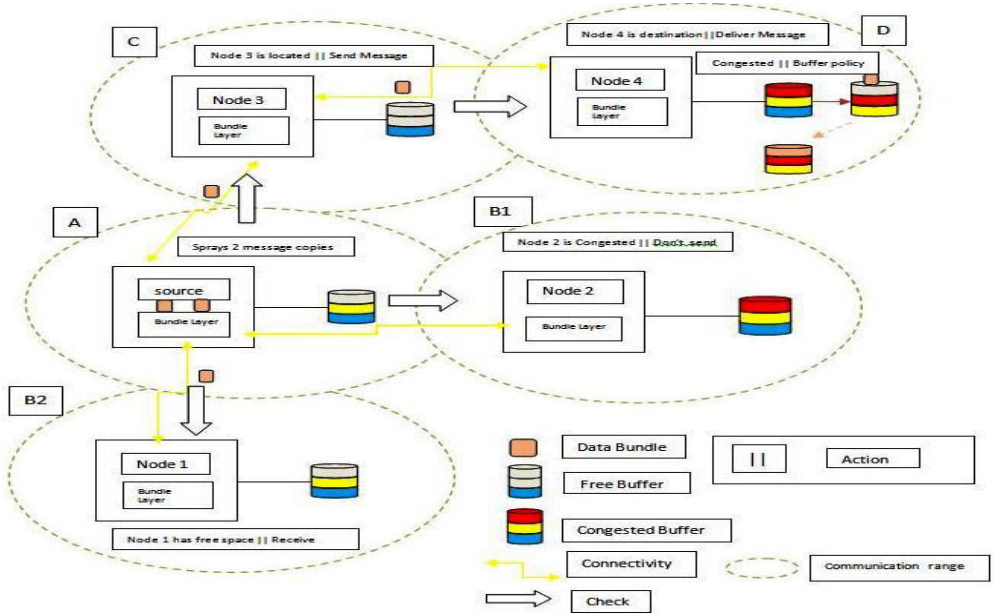

Figure 1.0 The Interaction of nodes in the VDTN using the CASaW protocol.

From the figure 1.0 above, Node A has 2 copies of message to be sprayed. Node A searches for nodes that are close in range and identifies nodes B1 and node B2. Before it sprays the message to node B1 and B2, it checks them to see that they have sufficient buffer space to receive the message. Node B2 has sufficient space to receive the message and a copy of the message is sent to it. Node B1 on the other hand, is congested as such Node A does not send the message to it because node B1 is congested and is not the destination node. Node A then searches for another node within its range and locates node C. Node C has sufficient buffer space and so receives the message. Node B2 and Node C are now to perform a direct delivery of the message to the destination. Node C identifies the destination (Node D) first and then checks to see if the destination has enough buffer space to receive the message. The destination node (D) is congested as such a buffer policy is applied whereby the oldest message in the buffer of the destination node is deleted to accommodate the new message. If node B2 later identifies the destination node (D) and sends the message to it, the destination node will reject the message because it already has a copy of the message.

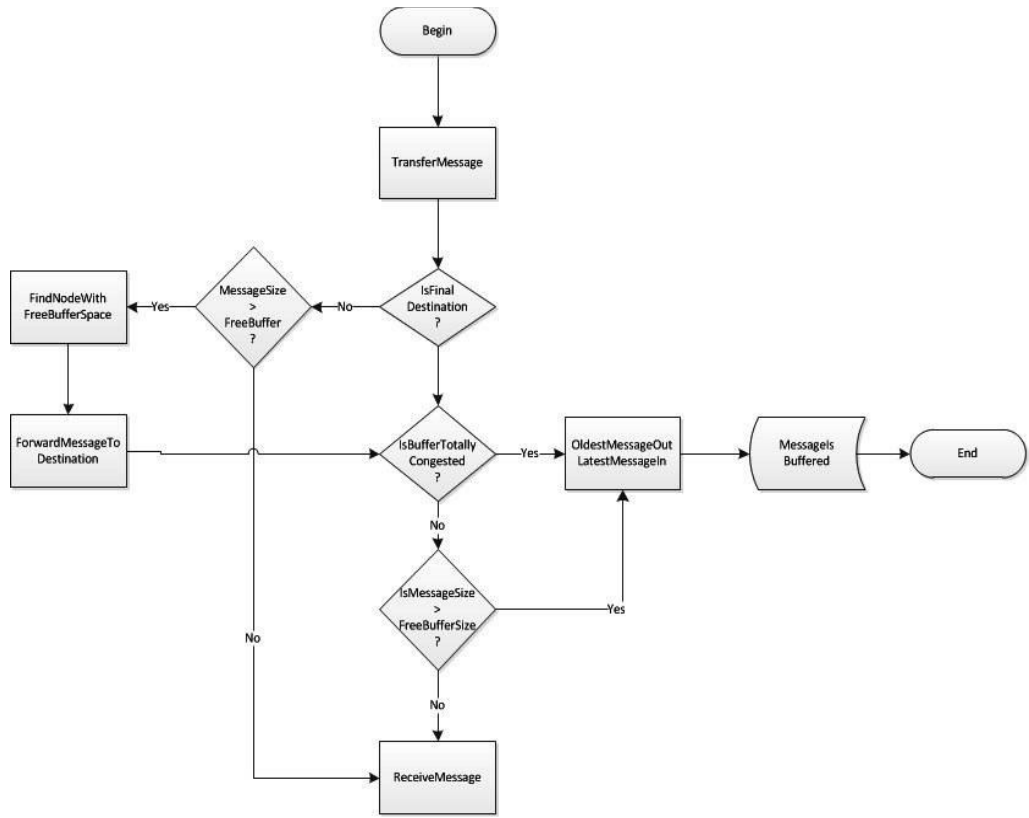

Figure 2.0 Message handling in the CASaW routing module.

## 5. EXPERIMENTS AND RESULTS

### A. Scenario Description

We have created 2 node groups with a total of 150 nodes. We implemented this amount of nodes to increase contact opportunities between nodes and to improve on node connectivity. Most specifically, we implemented the cars and the trams node groups. These groups have a general setting in common and they both have settings specific to them. For example, we have specified that the general number of nodes for the groups is 100 but for the tram group, we have specified 50 nodes. This means that group specific setting overrides the general settings for a specified parameter. The cars and the trams move along the Helsinki map paths on the roads and tram lines. The roads are specified for cars. The car group are configured to only drive on roads. They use the MapBasedMovement model which is the default setting for all node groups in the ONE simulator. The cars also have a speed range between (10 – 50) Km/h and have a wait time of (0 – 120) seconds. This means that cars can wait in the destination for (0 – 120) seconds. The tram group on the other hand on the roads through tram lines. They use the MapRouteMovement which has predefined routes and scheduled trips inside the city of Helsinki. Trams move at (7 – 10) Km/h speed and have a wait time of (10 – 30) seconds in routes defined for them.

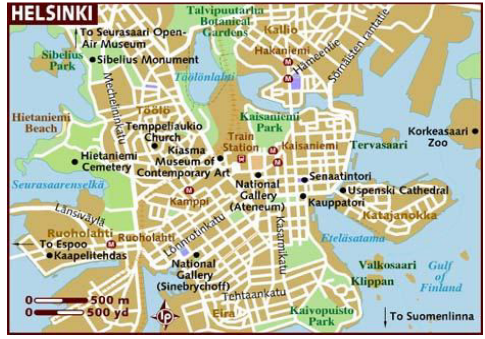

Figure 5.0: The Map of Helsinki, Finland used to simulate the behaviours of Nodes.

Node groups either use the blue-tooth interface or the high speed interface for communication. We focused on Vehicular Networks and our nodes are the Vehicles in the network which are characterized with high speed mobility. Therefore, we implemented the high speed interface which assumes a communication range of 100 meters at a transmission speed of 5Mbps because of its far reach compared to the blue-tooth interface which has a communication range of 10 meters and a transmission speed of 2Mbps. For our repetitive experiments, we specified a buffer size range of (5 – 45) MB so as to observe how increasing the size of the buffers impacts on nodal congestion. The size of the message has been set to 1MB. The message time-to-live (TTL) which is a mechanism that limits the amount of time of a message was set as 5 hours for a message. The simulation time was set to as 6 hours. This time excludes the warm-up time of 1000 seconds at the beginning of the simulation. The warm up time is specified so that nodes can be reasonably distributed over the Helsinki Map.

TABLE I Configuration of Experiments

| Parameters | Vehicles |
|---|---|
| Simulation time | 6 hours |
| Number of Nodes | 150 |
| Buffer Capacity | (5 – 45) MB |
| Interface | High Speed Interface |
| Speed | 5Mbps |
| Range | 100 meters |
| Mobility Patter | MapBasedMovement from Helsinki map |

**B. Metrics**

Two performance metrics have been considered to be most important to evaluate the impact of congestion in the VDTN. The performance metrics are: the number of message dropped and the number of message delivered. These performance measures are critical in the evaluation of the VDTN routing protocols and observing the impact of congestion in the Vehicular Delay Tolerant Networks (VDTN).

**C. Experiments**

We conducted a number of experiments. The configuration file of the ONE simulator was reconfigured for a significant number of times to fit specific simulation scenario. The goal of the experiment is to observe the changes in the number of messages dropped and the rate of message delivery as the size of the node buffer is increased. This is to observe the impact of congestion in the network and comparatively evaluate the VDTN routing protocols. To evaluate the impact of congestion in the network, it is considered very important to measure the amount of message drops in the network because message drops are indications to congestion in the network. We varied the buffer size of the nodes to gain an understanding of the impact of congestion in the network. The buffer sizes were varied from between 5MB to 45MB. We will discuss the result of our observation for each size of buffer below.

### 1.1. Results with 5MB Buffer Size

#### A. Messages Dropped

For this simulation scenario, we set the buffer of the nodes to be 5MB. After the running experiments for the VDTN routing protocols, we obtained the number of message dropped from the *MessageStats* file in the reports file of the ONE simulator. The table II below shows the amount of messages dropped when the buffer size is 5MB.

TABLE II Messages Dropped at 5MB Buffer Size

| Protocols | CASaW | SprayAndWait | MaxProp | Epidemic | Prophet |
|---|---|---|---|---|---|
| Message Dropped | 2064 | 3593 | 1699755 | 673920 | 1289574 |

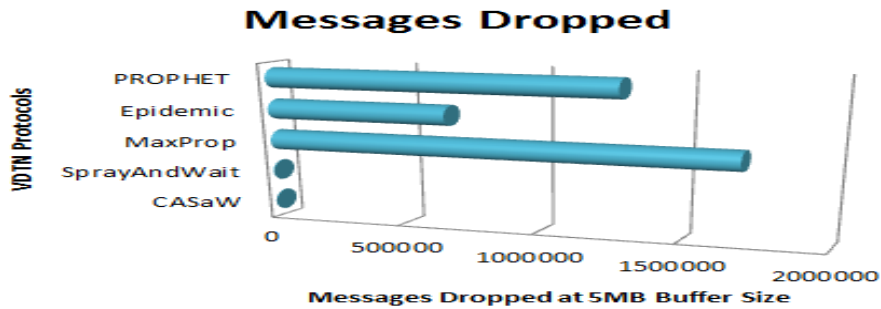

Figure 3.0 Messages dropped at 5MB Buffer Size

From the figure 3.0 above, it can be seen that at 5MB, our CASaW protocol dropped the least amount of message. The MaxProp protocol on the other hand dropped the highest amount of message followed by the Prophet and the Epidemic protocols respectively.

### 1.1. Results with 15MB Buffer Size

In this simulation scenario, we have increased the buffer size of the nodes to 15MB and then repeated the experiment for all routing VDTN protocols. The table 4.0 below shows the amount of messages dropped when the buffer size is 15MB

TABLE III Messages Dropped at 15MB Buffer Size

| Protocols | CASaW | SprayAndWait | MaxProp | Epidemic | Prophet |
|---|---|---|---|---|---|
| Message Dropped | 305 | 2103 | 26065 | 1755791 | 2120487 |

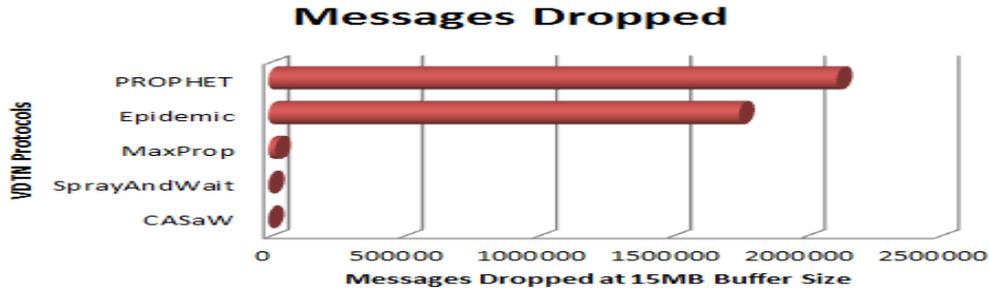

Figure 4.0 Messages dropped at 15MB Buffer Size

From the table III above it is seen that all evaluated VDTN routing protocol except the Epidemic and the Prophet protocol gained a reduction in the amount of message dropped. The CASaW routing protocol performed the least drop compared to all routing protocol followed by the benchmark protocol and then the MaxProp protocol which dropped the highest number of message when the buffer size was 5MB. The Epidemic routing protocol and the Prophet routing protocol which had the highest number of message drops had an increase in the amount of message dropped as the size of buffer increased.

## 1.1. Results with 25MB Buffer Size

The table IV below shows the number of messages dropped when the buffer is increased to 25MB.

TABLE IV Messages Dropped at 25MB Buffer Size

| Protocols | CASaW | SprayAndWait | MaxProp | Epidemic | Prophet |
| --- | --- | --- | --- | --- | --- |
| Message Dropped | 286 | 981 | 26065 | 2379605 | 2543794 |

From the table above, we see that there is a reduction in the number of message dropped for the Spray and Wait protocol, the MaxProp protocol and our CASaW protocol while the Epidemic protocol and the Prophet protocol gained an increase in the number of message dropped as the size of the node buffer increases from 15MB to 25MB. Our CASaW protocol achieved the least number of messages dropped compared to other evaluated VDTN routing protocol while the Prophet protocol achieved the highest number of message drops.

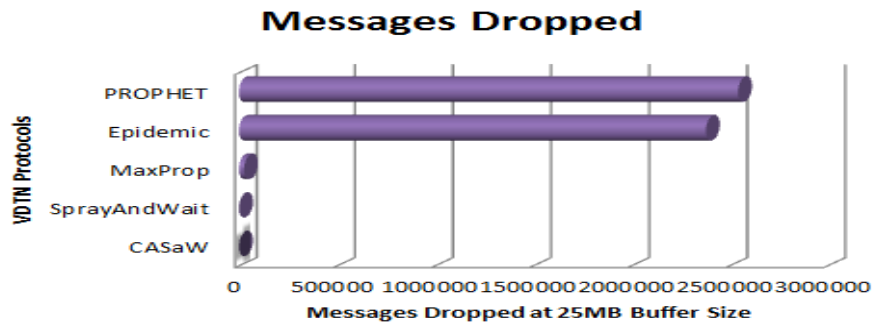

Figure 5.0 Messages dropped at 25MB Buffer Size

## 1.2. Results with 35MB Buffer Size

For this simulation scenario, we set the buffer of the nodes to be 35MB. After the running a series of simulations for the VDTN routing protocols, we obtained the number of message dropped as seen in the table below.

TABLE V Messages Dropped at 35MB Buffer Size

| Protocols | CASaW | SprayAndWait | MaxProp | Epidemic | Prophet |
|---|---|---|---|---|---|
| Message Dropped | 286 | 730 | 26065 | 2789483 | 2759765 |

From the table above, the CASaW and the MaxProp protocol presents a behaviour where no significant change is observed as the node buffer size is increased from 25MB to 35MB. One reason for this could be that the buffer size of the nodes becomes too big to contain messages and then messages are now dropped because of their time-to-live (TTL). This means that further increase in the size of the buffer of nodes could yield to same amount of drops.

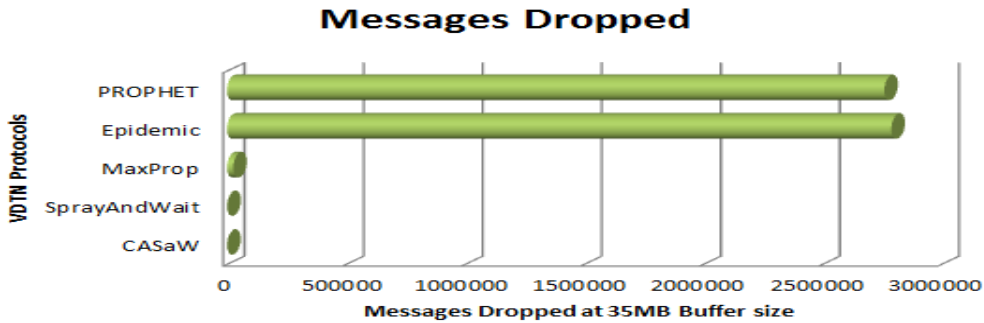

Figure 6.0 Messages dropped at 35MB Buffer Size

## 1.3. Results with 45MB Buffer Size

For this simulation scenario, we increased the buffer size of the nodes to 45MB. The table VI below shows the amount of messages dropped when the buffer size is 45MB.

TABLE VI Messages Dropped at 45MB Buffer Size

| Protocols | CASaW | SprayAndWait | MaxProp | Epidemic | Prophet |
|---|---|---|---|---|---|
| Message Dropped | 286 | 725 | 26065 | 3098227 | 2869954 |

From the table VI we see that the amount of message drops stays same for the CASaW and the MaxProp protocol when the buffer size of nodes is increased to 45MB. The Spray and Wait protocol had a reduction in the number of messages dropped while the Epidemic and the Prophet protocol had an increase in the amount of message dropped.

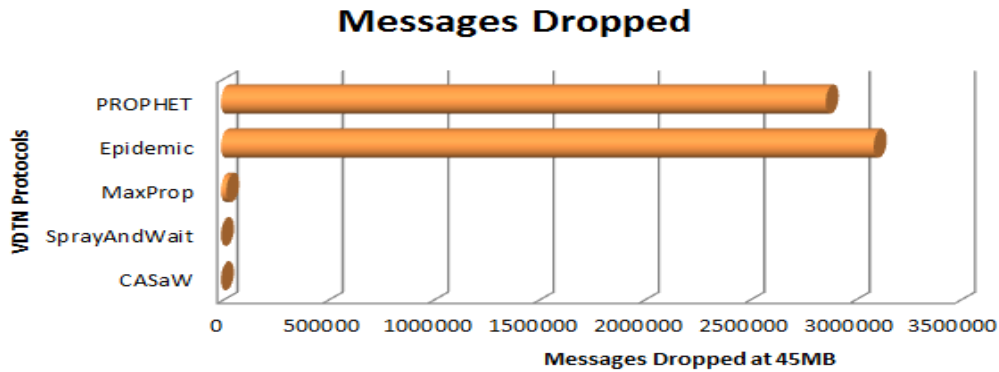

Figure 7.0 Messages dropped at 45MB Buffer Size

From the figure above, we see that the CASaW achieve the least amount of message drops compared to the Spray and Wait protocol and the MaxProp protocol. It also achieved the least amount of message drops when the buffer size of the nodes was increased to 45MB.

From the graphs, it can be said that compared to the benchmark protocol and other evaluated VDTN routing protocols, the CASaW protocol achieved the least number of message drop as the buffer size of the nodes in the network increases from 5MB to 45MB. It could therefore be interpreted to be that with the CASaW protocol, there will be minimal amount of congestion in the vehicular network and this could further mean that drivers in the vehicular environment could now be increasingly aware of their surroundings, road threats minimized and lives secured.

**B. Messages Delivered**

To evaluate the impact of congestion in the vehicular network, it is also important to measure the amount of messages delivered in a congestion prone vehicular environment. The buffer sizes of the nodes will vary from 5MB to 45MB. We will conduct experiments for each varied buffer size and observe the amount of messages delivered to the destination for all VDTN routing protocols.

The table VII below shows the number of message delivered for the VDTN routing protocols evaluated as the buffer size of the nodes increases from 5MB to 45MB.

TABLE VII  Messages Delivered as Buffer size increases from between 5MB to 45MB

| Buffer capacity | CASaW | SprayAndWait | MaxProp | Epidemic | Prophet |
| --- | --- | --- | --- | --- | --- |
| 5 MB | 665 | 677 | 543 | 80 | 137 |
| 15 MB | 651 | 698 | 720 | 117 | 167 |
| 25 MB | 648 | 699 | 720 | 149 | 183 |
| 35 MB | 648 | 699 | 720 | 170 | 195 |
| 45 MB | 648 | 699 | 720 | 209 | 203 |

From the table VII, when the buffer of the nodes is set at 5MB, the benchmark protocol performed more deliveries than our CASaW. The reason for this disparity might be because the benchmark protocol follows a short path to the destination and occurs in two phases. The Spray phase where messages are transferred to nearby nodes and the Wait phase where the located nodes perform a direct delivery to the destination if the destination node is not located at the spray phase. Our CASaW protocol on the other hand occurs in three phases. The check phase, the spray phase and the wait phase. It first check to see if the nearby nodes have free buffer space to contain the message before it sends the message and then, it sprays the message to only nodes with free space. If the destination is not located in the spray phase, the wait phase occurs. In this phase, the identified intermediate nodes with free spaces perform a direct delivery to the destination. In locating nodes with free buffer space, it follows a longer path to the destination therefore there we anticipate some amount of delay between message generation and message reception.

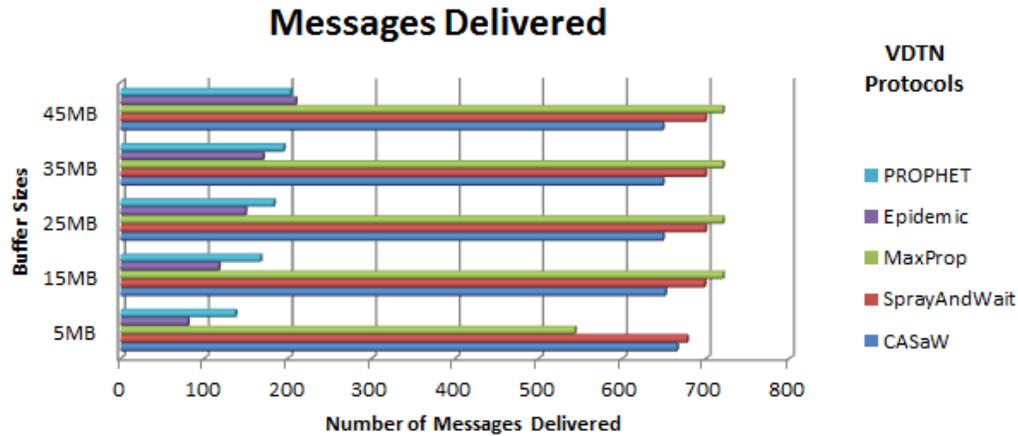

Figure 8.0 Messages delivered at varying buffer sizes

The CASaW performed more deliveries compared to other evaluated protocols i.e. the MaxProp protocol, the Epidemic protocol and the Prophet protocol when the buffer size of nodes is set at 5MB. The Epidemic protocol performed the least amount of deliveries.

When the buffer size is increased to 15MB, the MaxProp protocol delivered the highest number of messages to the destination compared to the benchmark protocol, our CASaW and the other evaluated protocols. The Epidemic protocol and the Prophet protocol also achieved an increase in the amount of message delivered as the buffer is increased to 15MB. The CASaW protocol which is an extension of the Spray and Wait protocol experienced a reduction in the amount of messages delivered. This is because the Spray and Wait and the CASaW protocol does not benefit from an increase in the size of the node buffer because of the limited amount of copies it sends [2] as such it impacts on the amount of messages it can actually deliver.

## 6. CONCLUSIONS

In this study, we considered congestion for the VDTN so as to minimize the rate of packet drops in the network in order to improve the awareness of drivers in the vehicular environment, optimize the probability of message delivery to the destination and improve road safety. To achieve this, we evaluated the generic VDTN routing protocols implemented in the ONE simulator so as to reliably make a choice of the most efficient routing protocol and use that protocol as our benchmark protocol. After evaluating the VDTN routing protocols, we selected the Spray and Wait protocol as our benchmark protocol because it performed better than other evaluated generic VDTN protocols. We then designed and implemented a congestion awareness algorithm for the spray and wait VDTN protocol so as to minimize the rate of packet losses in the vehicular network.

We conducted multiple experiments using the VDTN routing protocols. The experiments were conducted as the size of the buffer increases from 5MB to 15MB, 25MB, 35MB and 45MB. The results obtained showed that the CASaW protocol compared to other VDTN protocols performed the least amount of message drops as the size of the buffer of nodes in the network increases. It can therefore be said that the implementation of the CASaW in the VDTN would lead to a minimized rate of message drops and an increased awareness of drivers on the road. Also, a reduction in the amount of message dropped in the network can optimize the likelihood of message delivery to destinations.

We would like in the future to explore other possible sources of congestion in the VDTN and also like to evaluate the performance of the VDTN protocols using real data traces and other performance measures like latency and overhead ratio.